\documentclass[sigconf,screen]{acmart}

\AtBeginDocument{%
  \providecommand\BibTeX{{%
    \normalfont B\kern-0.5em{\scshape i\kern-0.25em b}\kern-0.8em\TeX}}}

\copyrightyear{2022}
\acmYear{2022}
\setcopyright{acmcopyright}\acmConference[ICPC '22]{30th International Conference on Program Comprehension}{May 16--17, 2022}{Virtual Event, USA}
\acmBooktitle{30th International Conference on Program Comprehension (ICPC '22), May 16--17, 2022, Virtual Event, USA}
\acmPrice{15.00}
\acmDOI{10.1145/3524610.3527876}
\acmISBN{978-1-4503-9298-3/22/05}

\newcommand{\squeezeup}{\vspace{-0.5cm}}
\usepackage{graphicx,calc}
\usepackage{hyperref}
\newlength\myheight
\newlength\mydepth
\settototalheight\myheight{Xygp}
\newcommand*\inlinegraphics[1]{%
  \settototalheight\myheight{Xygp}%
  \settodepth\mydepth{Xygp}%
  \raisebox{-\mydepth}{\includegraphics[height=\myheight]{#1}}%
}

\begin{document}

\title{GitQ- Towards Using Badges as Visual Cues for GitHub Projects}

\author{Akhila Sri Manasa Venigalla}
\authornote{Equal contribution}
\author{Kowndinya Boyalakuntla}
\authornotemark[1]
\author{Sridhar Chimalakonda}

\affiliation{%
  \institution{\textit{Research in Intelligent Software \& Human Analytics (RISHA) Lab}\\
  Department of Computer Science and Engineering\\
  Indian Institute of Technology Tirupati}
  \state{Tirupati}
  \country{India}
}
\email{{cs19d504, cs17b032, ch}@iittp.ac.in}

\begin{abstract}
  GitHub hosts millions of software repositories, facilitating developers to contribute to many projects in multiple ways. Most of the information about the repositories is text-based in the form of stars, forks, commits, and so on. However, developers willing to contribute to projects on GitHub often find it challenging to select appropriate projects to contribute to or reuse due to the large number of repositories present on GitHub. Further, obtaining this required information often becomes a tedious process, as one has to carefully mine information hidden inside the repository. To alleviate the effort intensive mining procedures, researchers have proposed npm-badges to outline information relating to build status of a project. However, these badges are static and limit their usage to package dependency and build details. \color{black} Adding visual cues such as badges, \inlinegraphics{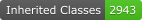}
 to the repositories might reduce the search space for developers. Hence, we present \textit{GitQ}, to automatically augment GitHub repositories with badges representing information about source code and project maintenance. Presenting \textit{GitQ} as a browser plugin to GitHub could make it easily accessible to developers using GitHub. \textit{GitQ} is evaluated with 15 developers based on the UTAUT model to understand developer perception towards its usefulness. We observed that 11 out of 15 developers perceived \textit{GitQ} to be useful in identifying the right set of repositories using visual cues such as \inlinegraphics{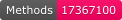} generated by \textit{GitQ}. The source code and tool are available for download on GitHub at \url{https://github.com/gitq-for-github/plugin}, and the demo can be found at \url{https://youtu.be/c0yohmIat3A}.
\end{abstract}



\maketitle
\pagestyle{plain}
\section{Introduction}

GitHub is extensively used among the available open-source code sharing platforms and currently hosts more than 72 million users\footnote{\url{https://github.com/search?q=type:user&type=Users}}, with nearly 43 million public repositories\footnote{\url{https://github.com/search?q=is:public}}. GitHub facilitates collaborative coding \cite{lima2014coding} and enables users to identify their peer developers' goals and suggest updates through pull requests, issues and commit messages \cite{dabbish2012social, liao2018exploring}. Most developers \textit{fork} a repository to reuse its source code if they find repositories with similar requirements \cite{gharehyazie2017some}. Readme-files and wikis of a repository are expected to provide information about the repository to the users \cite{zhang2017detecting, begel2013social}. However, it has been observed that the existing readme-files do not suffice to identify appropriate repositories \cite{kalliamvakou2014promises}. Even though it is effort-intensive, developers do not have a choice but to browse and analyze source code manually to understand about a repository. Researchers have developed several tools to assist developers in understanding GitHub repositories \cite{liao2018exploring, izquierdo2015gila, celinska2017programming}. These tools analyze tags on GitHub and provide visualizations \cite{izquierdo2015gila}, identify issue characteristics \cite{liao2018exploring}, support in understanding the network of programming languages used by users \cite{celinska2017programming}, and so on. 

Visualizing certain aspects of a project or repositories provide better comprehension about the project to developers \cite{merino2018towards}. Developers are facilitated with visualization of source code architecture \cite{gharibi2018code2graph}, the evolution of the project \cite{novais2013software} and so on.  
However, such visualizations are generally complicated and do not explicitly provide pointers about source code or maintenance. Providing such information might be of interest to novice developers who wish to contribute to repositories on GitHub \cite{borges2018s}. 

The usage of badges on crowd-sourced platforms had a positive impact on new users of the platform \cite{santos2020can}. Some repositories specific to Nodejs use \textit{npm-badges\footnote{\url{https://www.npmjs.com/package/badges}}} to represent a repository's characteristics, which are related to deployment and build details and hence provide a limited view of a repository specifically for the package called npm. 
Such badges are rarely valuable for novice developers who want to contribute to the repository based on the state of project maintenance, which could be useful for their onboarding.


\citet{legay2019usage} conducted a study on badges corresponding to build status, code coverage and so on, adopted by PHP and Rust package registries and observed that these badges could help developers in contributing to projects by instantly providing them details about a particular project across certain measures. 
Moreover, contributors in open-source projects volunteer their efforts and generally dedicate small amount of time to contribute and in the event of selecting poorly maintained projects, developer collaboration gets harder and such projects are likely to fail\cite{riehle2014paid, coelho2017modern}.

Therefore, using badges as visual cues could help GitHub users to identify projects of their choice, based on the maintenance and source code quality of the project.
\color{black}
Developer onboarding on GitHub has been studied in the literature to understand the factors that influence developer onboarding. Existing social relationships and prior language proficiency and experience were observed to be some of these factors \cite{casalnuovo2015developer}. NNLRank, a neural network model to suggest projects for better developer onboarding, based on developer skills and project features, has also been proposed \cite{liu2018recommending}. This network is based on a  ground-truth list of projects and considers only the developers' social profile and commit history of the projects. However, it does not consider the maintenance aspects of the project.

Hence, we propose \textit{GitQ}, as a tool towards adding badges such as \inlinegraphics{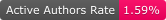}, \inlinegraphics{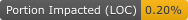} as visual cues to GitHub repositories. Such visual cues could help developers in understanding source code and maintenance metrics, without exploring repositories in detail. \textit{GitQ} displays metrics related to source code of the repository and maintenance status of the repository that could help in understanding the activeness and complexity of repositories, which are not explicitly present in the existing literature. We further evaluated \textit{GitQ} through a user survey, with 15 volunteers, using a 5-point Likert scale-based questionnaire adapted from the UTAUT technology acceptance model. Around 73\% of the participants found \textit{GitQ} to be useful and to be providing a good user experience.  Evaluation results and questionnaire can be found here\footnote{\url{https://drive.google.com/drive/folders/1Ldpo1gT-GDbv4TOGFH5GNwsO_aqD1zba?usp=sharing}}.

\section{Related Work}
GitHub provides various functionalities that help developers in developing, maintaining and extending their software. For instance, stargazers count, pull request lead time, programming languages used, act as popularity indicators and could be of interest to the contributors \cite{fronchetti2019attracts}. 
However, researchers have identified certain drawbacks on the existing GitHub platform such as difficulties in understanding commit messages, merged pull requests not being visible, unclear repository information, and so on \cite{kalliamvakou2014promises}, which can hinder the developer's productivity and cause a negative impact on user experience.
Visual cues such as badges have been implemented in crowd-sourced software platforms such as Stack Overflow, StackExchange \cite{santos2020can} and GitHub to enhance user experience \cite{trockman2018adding}.

\textbf{Visual Cues On Stack Overflow - }
Stack Overflow\footnote{\url{https://stackoverflow.com/}} provides badges to the questions based on the programming language used, the context of the problem, and so on. Venigalla et al. have proposed to tag questions on Stack Overflow based on their context and assign corresponding contextual badges \cite{venigalla2019sotagger}. 
Beyer et al. have also attempted to add badges on Stack Overflow based on the purpose of the questions on StackOverflow, which was further observed to be useful to Stack Overflow users \cite{beyer2018automatically}.

\textbf{Visual Cues On GitHub - }
A study on npm-badges assigned by project maintainers to repositories on GitHub revealed that badges such as build status, up-to-dateness, test coverage, and so on were primarily identified to be useful to contributors and users on GitHub \cite{trockman2018adding}. 
\textit{npm-badges} on GitHub repositories have facilitated researchers to analyze and obtain insights on types of continuous integration services, dependency managers adopted and underlying software tools used, across multiple repositories on GitHub \cite{lamba2020heard}. 
A study by Prana et al. revealed that most software professionals found the presence of automated labeling of GitHub content, in the form of badges, to be useful and to facilitate easy discovery of information \cite{prana2019categorizing}. 
Legay et al. have observed that most of the information presented by badges is static, highlighting the build status of the projects \cite{legay2019usage}.  

The npm-badges are to be manually added by the repository owners and are static in nature.
However, the deployment and build details provided through npm-badges do not provide maintenance related information such as the developer community in the repository, issue resolution rate and so on, which could be useful for developer onboarding.

Though badges or labels, automatically assigned, on crowd-sourced platforms such as StackExchange and Stack Overflow were observed to impact novice developers positively, usage of such automated visual cues is not yet explored on GitHub \cite{santos2020can, venigalla2019sotagger}. Automatically adding badges as visual cues to GitHub repositories could help novice developers gain better insights about the repositories, thus improving user experience on GitHub \cite{prana2019categorizing}. However, to the best of our knowledge, no tools or approaches have been presented to add automated badges based on source code and maintenance characteristics of GitHub repositories, other than for categorizing readme-files \cite{prana2019categorizing}.

Hence, we present \textit{GitQ}, a Google Chrome plugin, aimed to automatically annotate GitHub with badges based on metrics for source code\inlinegraphics{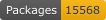} and maintenance\inlinegraphics{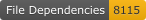}, generalized for a majority of GitHub repositories.

\section{Design and Development of \textit{GitQ}}
We designed \textit{GitQ} as a light-weight browser extension to convey essential information about a project. \textit{GitQ} quantifies the project's architecture and maintenance with few metrics as visual cue badges for the GitHub repositories, thereby assisting developers to gain insights. Metrics provide information about the repository that is not explicitly mentioned in the existing documentation, such as readme files, source code comments, and so on. \textit{GitQ} displays two categories of metrics - 1) source code metrics and 2) project maintenance metrics. 

We have defined the source code metrics to include dependencies among files in the repository, number of inherited classes, number of overridden methods, and so on in the repository across all files with the ".java" extension. These metrics are based on the commonly used CK Metrics \cite{chidamber1994metrics} defined to identify code quality for object-oriented languages. Considering the hierarchical organization of  repositories, CK metrics defined for object-oriented languages could align with the structure and hierarchy of GitHub repositories. The maintenance metrics include active authors rate, ratio of open and closed bug issues and extent of impact of commits on files and lines of code. These metrics could help in understanding the activeness of repository contributors, issue resolution rate in the repository and also average impact of commits in the repository. Detailed information about the insights shown by icons and definition of metrics are presented here\footnote{\url{https://drive.google.com/drive/folders/1xDnty_qAZr-gXsSMNQPxIpBOT5dZ8IZ2?usp=sharing}}.
Thus, \textit{GitQ} intends to portray some of the project's complex information as a visual feed to help developers better investigate the projects to contribute. 

\begin{figure}
    \centering
    \includegraphics[width=\linewidth]{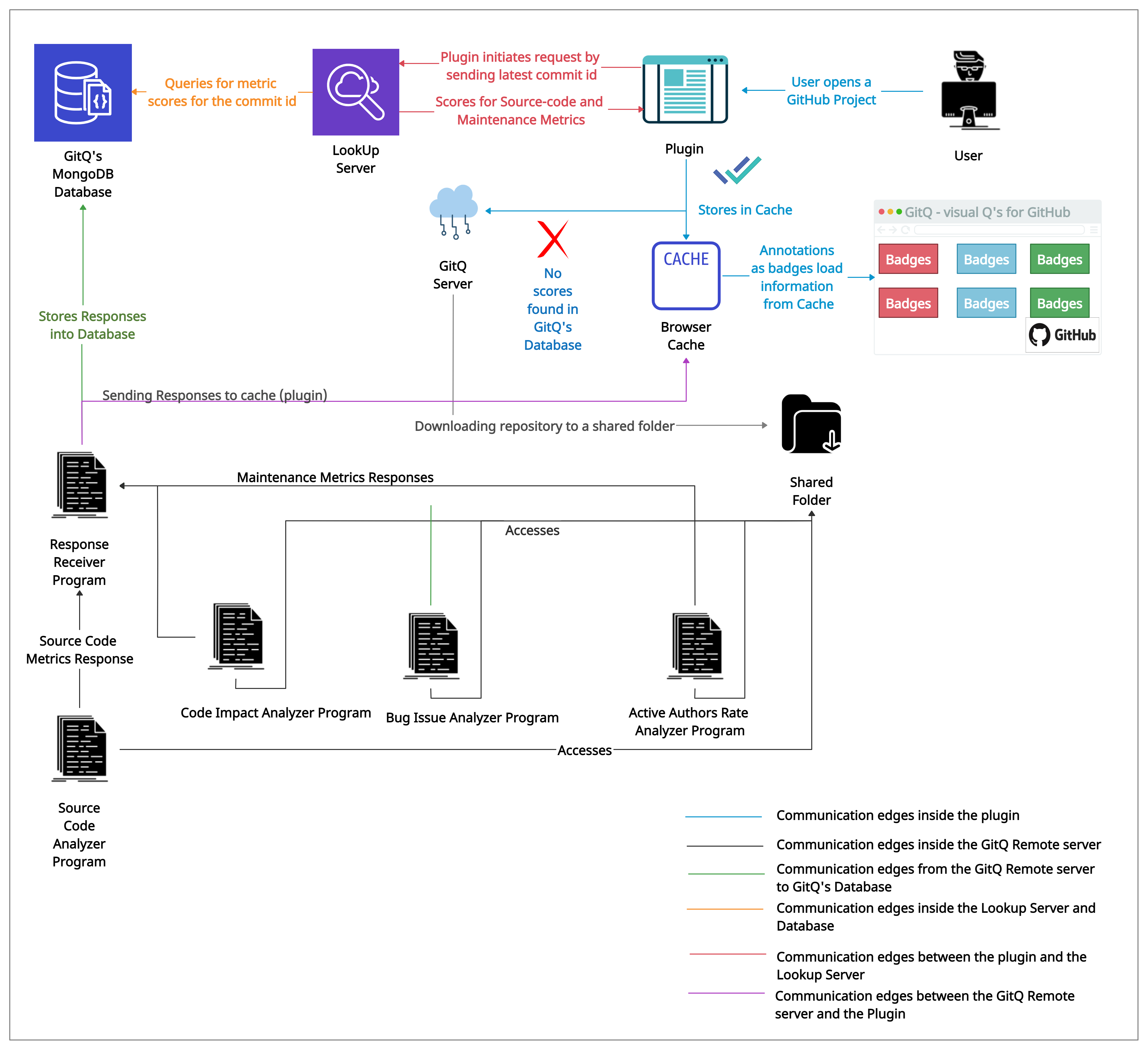}
    
  \captionof{figure}{Workflow diagram of \textit{GitQ}}
  \label{fig:overview}
  \squeezeup
\end{figure}

We implemented \textit{GitQ} as a browser extension, which displays badges on the GitHub repository page by sending the Project URL to a flask \footnote{\url{https://flask.palletsprojects.com/en/1.1.x/}}server. 
Figure \ref{fig:overview} depicts the work-flow of \textit{GitQ}. 
The back-end downloads the repository and then calculates values for the metrics proposed. It then sends the values back to the plugin, resting on the client's page. 
The starter program in the server receives a download request from the plugin. Metric calculations are routed at different URLs; hence metric calculations are executed in parallel. Python functions mapped to the routes access the shared folder (contains the downloaded repository) for processing metrics. Every response from the route is then sent back to the plugin as a JSON object. The repository is then automatically deleted. On the plugin side, the metric scores are stored inside the browser cache by mapping scores to commit id. Therefore, a project once analyzed, can be displayed with badges within no time for future visits until a new commit has been made to the repository.  

\section{User Scenario}
\begin{figure*}
    \centering
\includegraphics[scale=0.45]{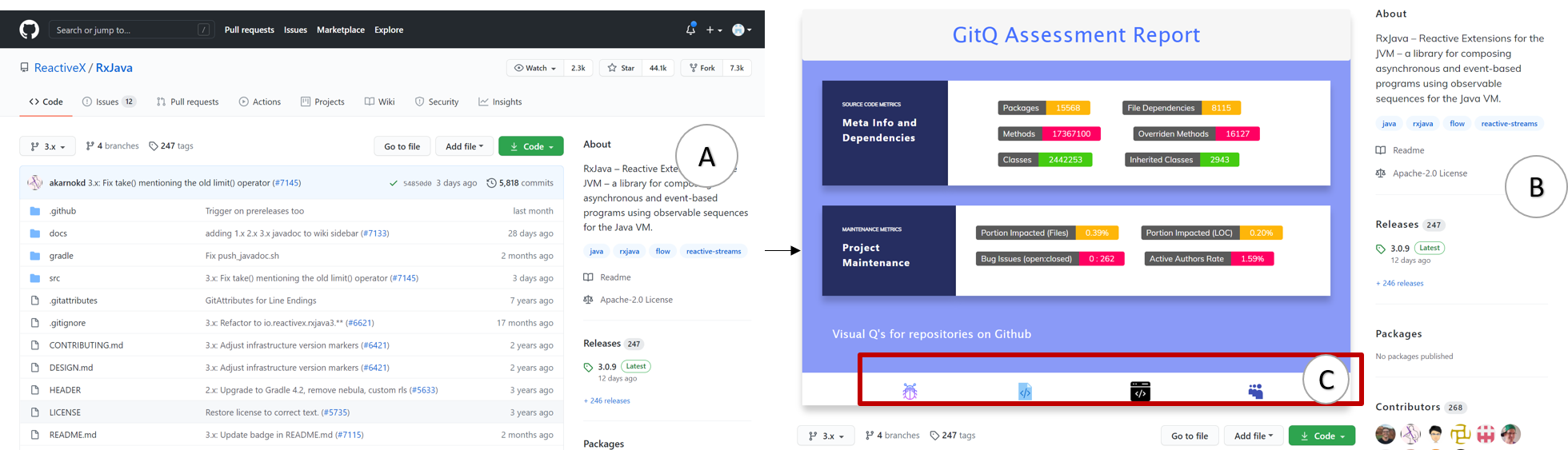}
\caption{User Scenario, with [A] depicting GitHub page prior to badges being added, [B] indicating the badges augmented and [C] indicating the list of insight-icons for maintenance metrics.}
    \label{fig:user_scenario}
    \squeezeup
\end{figure*}

Consider \textit{Moksha}, a novice developer, willing to contribute to projects on GitHub to enhance her project development skills. She intends to select projects that are well maintained, but confused about choosing the right project to contribute. She wishes to channel projects that are fairly maintained and well coded (less buggy) to get a glimpse of the project hierarchy and its maintenance over time. She discovers \textit{GitQ} as a tool that might reduce her effort in identifying the right repository. To use \textit{GitQ}, she loads the plugin into the browser and opens \textit{ReactiveX/RxJava}\footnote{\url{https://github.com/ReactiveX/RxJava}} (as depicted in Figure \ref{fig:user_scenario}[A]). \textit{GitQ} produces the report for the entire repository, as shown in Figure \ref{fig:user_scenario}[B]. She finds additional information about the project at a glance in the form of badges such as \inlinegraphics{gitq-badge-inherited-classes.png}, \inlinegraphics{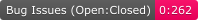}. For example, Moksha hovers onto Bugs icon (left-most icon in the Figure \ref{fig:user_scenario}[C]) and finds that the project has excellent bug handling but poor code composition. She discovers that at least 50\% of the code is affected for every commit. Moreover, she also observes that the project has a fairly inactive community (Less than 10\% of the authors remain active at any time); this would help her understand the activeness of a contributor in the project. Thus with \textit{GitQ}, Moksha gains better insights about the repository. This could help her escape the tedious process of manually browsing through the source code of the repository and its history.

\begin{figure}
    \centering
\includegraphics[scale=0.4]{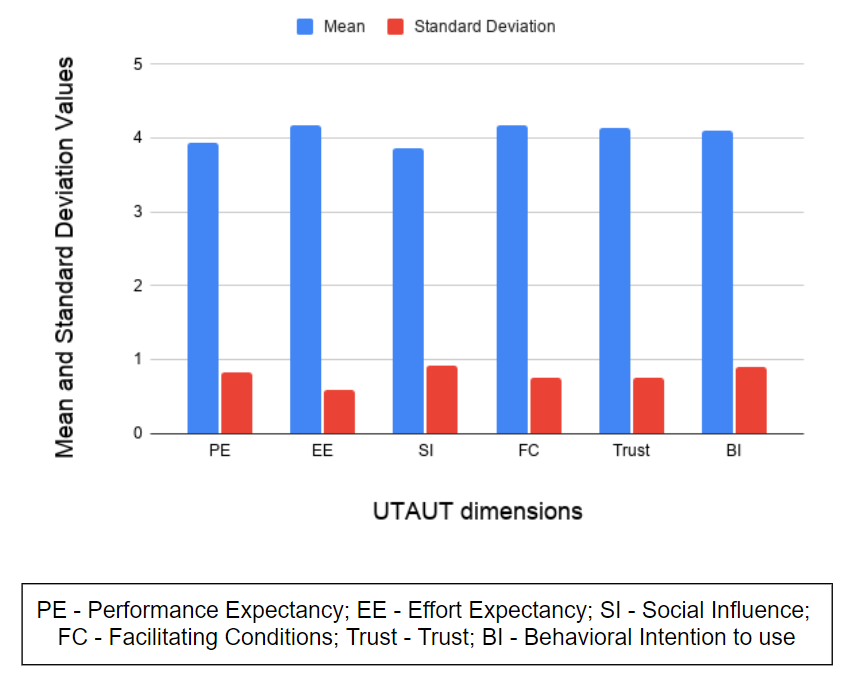}
\caption{Results of \textit{GitQ} User Survey}
    \label{fig:result}
    \squeezeup
\end{figure}

\section{Evaluation and Results}

\textit{GitQ} was developed to improve developers' experience while using GitHub and help them identify the right repository for reuse or contribution. Hence, we decided to evaluate \textit{GitQ} with respect to developers' perception towards its usability, usefulness, and correctness. 
The UTAUT (Unified Theory of Acceptance and Use of Technology) \cite{venkatesh2003user} model comprises of Performance Expectancy (PE), Effort Expectancy (EE), Social Influence (SI), Facilitating Conditions (FC) and Behavioural Intention to Use (BI) dimensions. Considering the need for \textit{GitQ} to be evaluated against similar dimensions, we chose to evaluate \textit{GitQ} using UTAUT model. 

We have added an additional dimension, Trust to fit the functionality of \textit{GitQ}. 
We conduct a pilot user survey\cite{legay2019usage, mcburney2014automatic} with 15 participants, to obtain insights on the user interpretation and usefulness of badge indication on a repository. We envision that obtaining insights on these aspects could be of potential use for future versions of \textit{GitQ}.
\color{black}
The 15 participant user survey comprised two research scholars, three undergraduate students, and ten professional developers working in the industry, with profound experience in using GitHub as a platform for open-source coding. 
\color{black}A document explaining the steps involved in installing and using the plugin is circulated among all the volunteers along with the survey form. The participants were requested to view at least 20 projects on GitHub with \textit{GitQ} plugin installed, to decide on the projects they wish to contribute, and to evaluate the correctness of the metrics used in \textit{GitQ}. The survey form comprised 12 questions based on the UTAUT questionnaire and five questions aimed towards understanding the volunteers' demographics. The survey participants were asked to fill in the Likert-scale based questionnaire based on their experiences after using \textit{GitQ}.

On an average the participants took around 10 minutes to view a repository and evaluate the correctness of metric values generated by \textit{GitQ}. The results of the questionnaire presented in Figure \ref{fig:result} indicate that majority of the volunteers agreed that \textit{GitQ} could be utilized as a indicator of project quality with the badges provided and might influence their decision in deciding GitHub projects \color{black} (Mean > 3.5 and Standard Deviation(SD) < 1). The mean values of the PE, EE, SI and FC are close to 4, indicating that \textit{GitQ} is easy to use with the existing technical expertise of the participants, and that \textit{GitQ} could help in obtaining insights about the project. It has also been observed that participants widely relied on the metrics generated by \textit{GitQ} (Mean = 4.13 and SD = 0.74), indicating a positive impact of the trust dimension.
The users have pointed out that projects with many files take a longer time to load and have also suggested extending \textit{GitQ} to display source code metrics for projects in other programming languages, apart from Java.


\section{Discussion and Limitations} 
%
%
%
%
Currently \textit{GitQ} is engineered in a way that the metric computation is handled solely by a remote server hosted at \url{https://services.iittp.ac.in}. Hiding the remote server, only the light-weight plugin is provided to the end user. However, the remote server might be inaccessible during downtime. Hence, we have decided to disentangle the server from the plugin and provide users with an alternative to use \textit{GitQ}. With this backup version of \textit{GitQ}\footnote{\url{https://github.com/gitq-for-github/plugin/tree/master/\%5Bstandalone-server-setup\%5Dbackup-version}}, users can run the metric computation on their own machines by following simple instructions outlined in the readme\footnote{\url{https://github.com/gitq-for-github/plugin/blob/master/\%5Bstandalone-server-setup\%5Dbackup-version/README.md}}. While the alternative version makes usability slightly difficult, it provides users with uninterrupted experience.

Currently, the scope for evaluating source code metrics is limited to java based projects, while maintenance metrics can be evaluated for any project in general. \textit{GitQ} could be extended to evaluate source code metrics for multiple programming languages. Also, the survey with 15 respondents might indicate biased results and might not generalize the opinion of a larger section of users. Thus, we plan to perform a robust survey, using an improved questionnaire, with a larger number of GitHub developers and contributors.  
\section{Conclusion and Future Work}
Developers coordinate their efforts into building projects on GitHub. To enhance the efficiency and user experience of developers, researchers have proposed various additions to the GitHub platform.
The automatic addition of badges to content on open source platforms such as Stack Overflow improves the developer's understanding of its content. Hence, we proposed \textit{GitQ}, as a Chrome plugin, to augment GitHub with badges that deliver visual cues quantifying the project's architecture and maintenance.  

The plugin forms the front-end of \textit{GitQ}, while the flask server in the back-end calculates values for the metrics chosen. Metrics and their values are shown as badges (\inlinegraphics{gitq-badge-inherited-classes.png}, \inlinegraphics{gitq-badge-packages.png}) on the screen. Source code metrics were identified to be in coherence with CK metrics. Apart from the badges, a summary of the project's maintenance is shown at the bottom of the assessment report for each of the maintenance metrics. From the developer survey, we found that majority of the volunteers agree (mean of PE=3.94) with \textit{GitQ} as a quick insight assistant to increase productivity and that they might adopt \textit{GitQ} while using GitHub for their project development \color{black}. They also agree to recommend \textit{GitQ} to their peers. 

As our goal is to propose the idea of using badges as quality indicator for GitHub projects, we demonstrated use of badges using a few metrics. The metrics currently present in \textit{GitQ} are an effort to signal underlying project quality and maintenance. However, there is a tremendous scope to include metrics that better qualify for the definition of a well maintained or a good quality project.  Hence, as a part of the future work, we intend to include more metrics that can identify the underlying best-practices of open-source projects. We also intend to display trends (in the form of graphs) observed in open-source projects to motivate developers to follow best practices.  

\balance
\bibliographystyle{ACM-Reference-Format}
\bibliography{sample-base}


\begin{thebibliography}{28}


\ifx \showCODEN    \undefined \def \showCODEN     #1{\unskip}     \fi
\ifx \showDOI      \undefined \def \showDOI       #1{#1}\fi
\ifx \showISBNx    \undefined \def \showISBNx     #1{\unskip}     \fi
\ifx \showISBNxiii \undefined \def \showISBNxiii  #1{\unskip}     \fi
\ifx \showISSN     \undefined \def \showISSN      #1{\unskip}     \fi
\ifx \showLCCN     \undefined \def \showLCCN      #1{\unskip}     \fi
\ifx \shownote     \undefined \def \shownote      #1{#1}          \fi
\ifx \showarticletitle \undefined \def \showarticletitle #1{#1}   \fi
\ifx \showURL      \undefined \def \showURL       {\relax}        \fi
\providecommand\bibfield[2]{#2}
\providecommand\bibinfo[2]{#2}
\providecommand\natexlab[1]{#1}
\providecommand\showeprint[2][]{arXiv:#2}

\bibitem[\protect\citeauthoryear{Begel, Bosch, and Storey}{Begel
  et~al\mbox{.}}{2013}]%
        {begel2013social}
\bibfield{author}{\bibinfo{person}{Andrew Begel}, \bibinfo{person}{Jan Bosch},
  {and} \bibinfo{person}{Margaret-Anne Storey}.}
  \bibinfo{year}{2013}\natexlab{}.
\newblock \showarticletitle{Social networking meets software development:
  Perspectives from github, msdn, stack exchange, and topcoder}.
\newblock \bibinfo{journal}{\emph{IEEE Software}} \bibinfo{volume}{30},
  \bibinfo{number}{1} (\bibinfo{year}{2013}), \bibinfo{pages}{52--66}.
\newblock


\bibitem[\protect\citeauthoryear{Beyer, Macho, Di~Penta, and Pinzger}{Beyer
  et~al\mbox{.}}{2018}]%
        {beyer2018automatically}
\bibfield{author}{\bibinfo{person}{Stefanie Beyer}, \bibinfo{person}{Christian
  Macho}, \bibinfo{person}{Massimiliano Di~Penta}, {and}
  \bibinfo{person}{Martin Pinzger}.} \bibinfo{year}{2018}\natexlab{}.
\newblock \showarticletitle{Automatically classifying posts into question
  categories on stack overflow}. In \bibinfo{booktitle}{\emph{2018 IEEE/ACM
  26th International Conference on Program Comprehension (ICPC)}}. IEEE,
  \bibinfo{pages}{211--21110}.
\newblock


\bibitem[\protect\citeauthoryear{Borges and Valente}{Borges and
  Valente}{2018}]%
        {borges2018s}
\bibfield{author}{\bibinfo{person}{Hudson Borges} {and}
  \bibinfo{person}{Marco~Tulio Valente}.} \bibinfo{year}{2018}\natexlab{}.
\newblock \showarticletitle{What’s in a GitHub star? understanding repository
  starring practices in a social coding platform}.
\newblock \bibinfo{journal}{\emph{Journal of Systems and Software}}
  \bibinfo{volume}{146} (\bibinfo{year}{2018}), \bibinfo{pages}{112--129}.
\newblock


\bibitem[\protect\citeauthoryear{Casalnuovo, Vasilescu, Devanbu, and
  Filkov}{Casalnuovo et~al\mbox{.}}{2015}]%
        {casalnuovo2015developer}
\bibfield{author}{\bibinfo{person}{Casey Casalnuovo}, \bibinfo{person}{Bogdan
  Vasilescu}, \bibinfo{person}{Premkumar Devanbu}, {and}
  \bibinfo{person}{Vladimir Filkov}.} \bibinfo{year}{2015}\natexlab{}.
\newblock \showarticletitle{Developer onboarding in GitHub: the role of prior
  social links and language experience}. In
  \bibinfo{booktitle}{\emph{Proceedings of the 2015 10th joint meeting on
  foundations of software engineering}}. \bibinfo{pages}{817--828}.
\newblock


\bibitem[\protect\citeauthoryear{Celi{\'n}ska and Kopczy{\'n}ski}{Celi{\'n}ska
  and Kopczy{\'n}ski}{2017}]%
        {celinska2017programming}
\bibfield{author}{\bibinfo{person}{Dorota Celi{\'n}ska} {and}
  \bibinfo{person}{Eryk Kopczy{\'n}ski}.} \bibinfo{year}{2017}\natexlab{}.
\newblock \showarticletitle{Programming languages in github: a visualization in
  hyperbolic plane}. In \bibinfo{booktitle}{\emph{Eleventh International AAAI
  Conference on Web and Social Media}}.
\newblock


\bibitem[\protect\citeauthoryear{Chidamber and Kemerer}{Chidamber and
  Kemerer}{1994}]%
        {chidamber1994metrics}
\bibfield{author}{\bibinfo{person}{Shyam~R Chidamber} {and}
  \bibinfo{person}{Chris~F Kemerer}.} \bibinfo{year}{1994}\natexlab{}.
\newblock \showarticletitle{A metrics suite for object oriented design}.
\newblock \bibinfo{journal}{\emph{IEEE Transactions on software engineering}}
  \bibinfo{volume}{20}, \bibinfo{number}{6} (\bibinfo{year}{1994}),
  \bibinfo{pages}{476--493}.
\newblock


\bibitem[\protect\citeauthoryear{Coelho and Valente}{Coelho and
  Valente}{2017}]%
        {coelho2017modern}
\bibfield{author}{\bibinfo{person}{Jailton Coelho} {and}
  \bibinfo{person}{Marco~Tulio Valente}.} \bibinfo{year}{2017}\natexlab{}.
\newblock \showarticletitle{Why modern open source projects fail}. In
  \bibinfo{booktitle}{\emph{Proceedings of the 2017 11th Joint Meeting on
  Foundations of Software Engineering}}. \bibinfo{pages}{186--196}.
\newblock


\bibitem[\protect\citeauthoryear{Dabbish, Stuart, Tsay, and Herbsleb}{Dabbish
  et~al\mbox{.}}{2012}]%
        {dabbish2012social}
\bibfield{author}{\bibinfo{person}{Laura Dabbish}, \bibinfo{person}{Colleen
  Stuart}, \bibinfo{person}{Jason Tsay}, {and} \bibinfo{person}{Jim Herbsleb}.}
  \bibinfo{year}{2012}\natexlab{}.
\newblock \showarticletitle{Social coding in GitHub: transparency and
  collaboration in an open software repository}. In
  \bibinfo{booktitle}{\emph{Proceedings of the ACM 2012 conference on computer
  supported cooperative work}}. ACM, \bibinfo{pages}{1277--1286}.
\newblock


\bibitem[\protect\citeauthoryear{Fronchetti, Wiese, Pinto, and
  Steinmacher}{Fronchetti et~al\mbox{.}}{2019}]%
        {fronchetti2019attracts}
\bibfield{author}{\bibinfo{person}{Felipe Fronchetti}, \bibinfo{person}{Igor
  Wiese}, \bibinfo{person}{Gustavo Pinto}, {and} \bibinfo{person}{Igor
  Steinmacher}.} \bibinfo{year}{2019}\natexlab{}.
\newblock \showarticletitle{What attracts newcomers to onboard on oss projects?
  tl; dr: Popularity}. In \bibinfo{booktitle}{\emph{IFIP International
  Conference on Open Source Systems}}. Springer, \bibinfo{pages}{91--103}.
\newblock


\bibitem[\protect\citeauthoryear{Gharehyazie, Ray, and Filkov}{Gharehyazie
  et~al\mbox{.}}{2017}]%
        {gharehyazie2017some}
\bibfield{author}{\bibinfo{person}{Mohammad Gharehyazie},
  \bibinfo{person}{Baishakhi Ray}, {and} \bibinfo{person}{Vladimir Filkov}.}
  \bibinfo{year}{2017}\natexlab{}.
\newblock \showarticletitle{Some from here, some from there: Cross-project code
  reuse in github}. In \bibinfo{booktitle}{\emph{2017 IEEE/ACM 14th
  International Conference on Mining Software Repositories (MSR)}}. IEEE,
  \bibinfo{pages}{291--301}.
\newblock


\bibitem[\protect\citeauthoryear{Gharibi, Tripathi, and Lee}{Gharibi
  et~al\mbox{.}}{2018}]%
        {gharibi2018code2graph}
\bibfield{author}{\bibinfo{person}{Gharib Gharibi}, \bibinfo{person}{Rashmi
  Tripathi}, {and} \bibinfo{person}{Yugyung Lee}.}
  \bibinfo{year}{2018}\natexlab{}.
\newblock \showarticletitle{Code2graph: automatic generation of static call
  graphs for python source code}. In \bibinfo{booktitle}{\emph{Proceedings of
  the 33rd ACM/IEEE International Conference on Automated Software
  Engineering}}. \bibinfo{pages}{880--883}.
\newblock


\bibitem[\protect\citeauthoryear{Izquierdo, Cosentino, Rolandi, Bergel, and
  Cabot}{Izquierdo et~al\mbox{.}}{2015}]%
        {izquierdo2015gila}
\bibfield{author}{\bibinfo{person}{Javier Luis~C{\'a}novas Izquierdo},
  \bibinfo{person}{Valerio Cosentino}, \bibinfo{person}{Bel{\'e}n Rolandi},
  \bibinfo{person}{Alexandre Bergel}, {and} \bibinfo{person}{Jordi Cabot}.}
  \bibinfo{year}{2015}\natexlab{}.
\newblock \showarticletitle{Gila: Github label analyzer}. In
  \bibinfo{booktitle}{\emph{2015 IEEE 22nd International Conference on Software
  Analysis, Evolution, and Reengineering (SANER)}}. IEEE,
  \bibinfo{pages}{479--483}.
\newblock


\bibitem[\protect\citeauthoryear{Kalliamvakou, Gousios, Blincoe, Singer,
  German, and Damian}{Kalliamvakou et~al\mbox{.}}{2014}]%
        {kalliamvakou2014promises}
\bibfield{author}{\bibinfo{person}{Eirini Kalliamvakou},
  \bibinfo{person}{Georgios Gousios}, \bibinfo{person}{Kelly Blincoe},
  \bibinfo{person}{Leif Singer}, \bibinfo{person}{Daniel~M German}, {and}
  \bibinfo{person}{Daniela Damian}.} \bibinfo{year}{2014}\natexlab{}.
\newblock \showarticletitle{The promises and perils of mining GitHub}. In
  \bibinfo{booktitle}{\emph{Proceedings of the 11th working conference on
  mining software repositories}}. ACM, \bibinfo{pages}{92--101}.
\newblock


\bibitem[\protect\citeauthoryear{Lamba, Trockman, Armanios, K{\"a}stner,
  Miller, and Vasilescu}{Lamba et~al\mbox{.}}{2020}]%
        {lamba2020heard}
\bibfield{author}{\bibinfo{person}{Hemank Lamba}, \bibinfo{person}{Asher
  Trockman}, \bibinfo{person}{Daniel Armanios}, \bibinfo{person}{Christian
  K{\"a}stner}, \bibinfo{person}{Heather Miller}, {and} \bibinfo{person}{Bogdan
  Vasilescu}.} \bibinfo{year}{2020}\natexlab{}.
\newblock \showarticletitle{Heard it through the Gitvine: an empirical study of
  tool diffusion across the npm ecosystem}. In
  \bibinfo{booktitle}{\emph{Proceedings of the 28th ACM Joint Meeting on
  European Software Engineering Conference and Symposium on the Foundations of
  Software Engineering}}. \bibinfo{pages}{505--517}.
\newblock


\bibitem[\protect\citeauthoryear{Legay, Decan, and Mens}{Legay
  et~al\mbox{.}}{2019}]%
        {legay2019usage}
\bibfield{author}{\bibinfo{person}{Damien Legay}, \bibinfo{person}{Alexandre
  Decan}, {and} \bibinfo{person}{Tom Mens}.} \bibinfo{year}{2019}\natexlab{}.
\newblock \showarticletitle{On the Usage of Badges in Open Source Packages on
  GitHub.}. In \bibinfo{booktitle}{\emph{BENEVOL}}.
\newblock


\bibitem[\protect\citeauthoryear{Liao, He, Chen, Fan, Zhang, and Liu}{Liao
  et~al\mbox{.}}{2018}]%
        {liao2018exploring}
\bibfield{author}{\bibinfo{person}{Zhifang Liao}, \bibinfo{person}{Dayu He},
  \bibinfo{person}{Zhijie Chen}, \bibinfo{person}{Xiaoping Fan},
  \bibinfo{person}{Yan Zhang}, {and} \bibinfo{person}{Shengzong Liu}.}
  \bibinfo{year}{2018}\natexlab{}.
\newblock \showarticletitle{Exploring the characteristics of issue-related
  behaviors in GitHub using visualization techniques}.
\newblock \bibinfo{journal}{\emph{IEEE Access}}  \bibinfo{volume}{6}
  (\bibinfo{year}{2018}), \bibinfo{pages}{24003--24015}.
\newblock


\bibitem[\protect\citeauthoryear{Lima, Rossi, and Musolesi}{Lima
  et~al\mbox{.}}{2014}]%
        {lima2014coding}
\bibfield{author}{\bibinfo{person}{Antonio Lima}, \bibinfo{person}{Luca Rossi},
  {and} \bibinfo{person}{Mirco Musolesi}.} \bibinfo{year}{2014}\natexlab{}.
\newblock \showarticletitle{Coding together at scale: GitHub as a collaborative
  social network}. In \bibinfo{booktitle}{\emph{Eighth International AAAI
  Conference on Weblogs and Social Media}}.
\newblock


\bibitem[\protect\citeauthoryear{Liu, Yang, Zhang, Ray, and Rahman}{Liu
  et~al\mbox{.}}{2018}]%
        {liu2018recommending}
\bibfield{author}{\bibinfo{person}{Chao Liu}, \bibinfo{person}{Dan Yang},
  \bibinfo{person}{Xiaohong Zhang}, \bibinfo{person}{Baishakhi Ray}, {and}
  \bibinfo{person}{Md~Masudur Rahman}.} \bibinfo{year}{2018}\natexlab{}.
\newblock \showarticletitle{Recommending github projects for developer
  onboarding}.
\newblock \bibinfo{journal}{\emph{IEEE Access}}  \bibinfo{volume}{6}
  (\bibinfo{year}{2018}), \bibinfo{pages}{52082--52094}.
\newblock


\bibitem[\protect\citeauthoryear{McBurney and McMillan}{McBurney and
  McMillan}{2014}]%
        {mcburney2014automatic}
\bibfield{author}{\bibinfo{person}{Paul~W McBurney} {and}
  \bibinfo{person}{Collin McMillan}.} \bibinfo{year}{2014}\natexlab{}.
\newblock \showarticletitle{Automatic documentation generation via source code
  summarization of method context}. In \bibinfo{booktitle}{\emph{Proceedings of
  the 22nd International Conference on Program Comprehension}}.
  \bibinfo{pages}{279--290}.
\newblock


\bibitem[\protect\citeauthoryear{Merino, Ghafari, and Nierstrasz}{Merino
  et~al\mbox{.}}{2018}]%
        {merino2018towards}
\bibfield{author}{\bibinfo{person}{Leonel Merino}, \bibinfo{person}{Mohammad
  Ghafari}, {and} \bibinfo{person}{Oscar Nierstrasz}.}
  \bibinfo{year}{2018}\natexlab{}.
\newblock \showarticletitle{Towards actionable visualization for software
  developers}.
\newblock \bibinfo{journal}{\emph{Journal of software: evolution and process}}
  \bibinfo{volume}{30}, \bibinfo{number}{2} (\bibinfo{year}{2018}),
  \bibinfo{pages}{e1923}.
\newblock


\bibitem[\protect\citeauthoryear{Novais, Torres, Mendes, Mendon{\c{c}}a, and
  Zazworka}{Novais et~al\mbox{.}}{2013}]%
        {novais2013software}
\bibfield{author}{\bibinfo{person}{Renato~Lima Novais},
  \bibinfo{person}{Andr{\'e} Torres}, \bibinfo{person}{Thiago~Souto Mendes},
  \bibinfo{person}{Manoel Mendon{\c{c}}a}, {and} \bibinfo{person}{Nico
  Zazworka}.} \bibinfo{year}{2013}\natexlab{}.
\newblock \showarticletitle{Software evolution visualization: A systematic
  mapping study}.
\newblock \bibinfo{journal}{\emph{Information and Software Technology}}
  \bibinfo{volume}{55}, \bibinfo{number}{11} (\bibinfo{year}{2013}),
  \bibinfo{pages}{1860--1883}.
\newblock


\bibitem[\protect\citeauthoryear{Prana, Treude, Thung, Atapattu, and Lo}{Prana
  et~al\mbox{.}}{2019}]%
        {prana2019categorizing}
\bibfield{author}{\bibinfo{person}{Gede Artha~Azriadi Prana},
  \bibinfo{person}{Christoph Treude}, \bibinfo{person}{Ferdian Thung},
  \bibinfo{person}{Thushari Atapattu}, {and} \bibinfo{person}{David Lo}.}
  \bibinfo{year}{2019}\natexlab{}.
\newblock \showarticletitle{Categorizing the content of GitHub README files}.
\newblock \bibinfo{journal}{\emph{Empirical Software Engineering}}
  \bibinfo{volume}{24}, \bibinfo{number}{3} (\bibinfo{year}{2019}),
  \bibinfo{pages}{1296--1327}.
\newblock


\bibitem[\protect\citeauthoryear{Riehle, Riemer, Kolassa, and Schmidt}{Riehle
  et~al\mbox{.}}{2014}]%
        {riehle2014paid}
\bibfield{author}{\bibinfo{person}{Dirk Riehle}, \bibinfo{person}{Philipp
  Riemer}, \bibinfo{person}{Carsten Kolassa}, {and} \bibinfo{person}{Michael
  Schmidt}.} \bibinfo{year}{2014}\natexlab{}.
\newblock \showarticletitle{Paid vs. volunteer work in open source}. In
  \bibinfo{booktitle}{\emph{2014 47th Hawaii International Conference on System
  Sciences}}. IEEE, \bibinfo{pages}{3286--3295}.
\newblock


\bibitem[\protect\citeauthoryear{Santos, Burghardt, Lerman, and Helic}{Santos
  et~al\mbox{.}}{2020}]%
        {santos2020can}
\bibfield{author}{\bibinfo{person}{Tiago Santos}, \bibinfo{person}{Keith
  Burghardt}, \bibinfo{person}{Kristina Lerman}, {and} \bibinfo{person}{Denis
  Helic}.} \bibinfo{year}{2020}\natexlab{}.
\newblock \showarticletitle{Can Badges Foster a More Welcoming Culture on Q\&A
  Boards?}. In \bibinfo{booktitle}{\emph{Proceedings of the International AAAI
  Conference on Web and Social Media}}, Vol.~\bibinfo{volume}{14}.
  \bibinfo{pages}{969--973}.
\newblock


\bibitem[\protect\citeauthoryear{Trockman, Zhou, K{\"a}stner, and
  Vasilescu}{Trockman et~al\mbox{.}}{2018}]%
        {trockman2018adding}
\bibfield{author}{\bibinfo{person}{Asher Trockman}, \bibinfo{person}{Shurui
  Zhou}, \bibinfo{person}{Christian K{\"a}stner}, {and} \bibinfo{person}{Bogdan
  Vasilescu}.} \bibinfo{year}{2018}\natexlab{}.
\newblock \showarticletitle{Adding sparkle to social coding: an empirical study
  of repository badges in the npm ecosystem}. In
  \bibinfo{booktitle}{\emph{Proceedings of the 40th International Conference on
  Software Engineering}}. \bibinfo{pages}{511--522}.
\newblock


\bibitem[\protect\citeauthoryear{Venigalla, Lakkundi, and
  Chimalakonda}{Venigalla et~al\mbox{.}}{2019}]%
        {venigalla2019sotagger}
\bibfield{author}{\bibinfo{person}{Akhila Sri~Manasa Venigalla},
  \bibinfo{person}{Chaitanya~S Lakkundi}, {and} \bibinfo{person}{Sridhar
  Chimalakonda}.} \bibinfo{year}{2019}\natexlab{}.
\newblock \showarticletitle{SOTagger-Towards Classifying Stack Overflow Posts
  through Contextual Tagging (S).}. In \bibinfo{booktitle}{\emph{SEKE}}.
  \bibinfo{pages}{493--639}.
\newblock


\bibitem[\protect\citeauthoryear{Venkatesh, Morris, Davis, and Davis}{Venkatesh
  et~al\mbox{.}}{2003}]%
        {venkatesh2003user}
\bibfield{author}{\bibinfo{person}{Viswanath Venkatesh},
  \bibinfo{person}{Michael~G Morris}, \bibinfo{person}{Gordon~B Davis}, {and}
  \bibinfo{person}{Fred~D Davis}.} \bibinfo{year}{2003}\natexlab{}.
\newblock \showarticletitle{User acceptance of information technology: Toward a
  unified view}.
\newblock \bibinfo{journal}{\emph{MIS quarterly}} (\bibinfo{year}{2003}),
  \bibinfo{pages}{425--478}.
\newblock


\bibitem[\protect\citeauthoryear{Zhang, Lo, Kochhar, Xia, Li, and Sun}{Zhang
  et~al\mbox{.}}{2017}]%
        {zhang2017detecting}
\bibfield{author}{\bibinfo{person}{Yun Zhang}, \bibinfo{person}{David Lo},
  \bibinfo{person}{Pavneet~Singh Kochhar}, \bibinfo{person}{Xin Xia},
  \bibinfo{person}{Quanlai Li}, {and} \bibinfo{person}{Jianling Sun}.}
  \bibinfo{year}{2017}\natexlab{}.
\newblock \showarticletitle{Detecting similar repositories on GitHub}. In
  \bibinfo{booktitle}{\emph{2017 IEEE 24th International Conference on Software
  Analysis, Evolution and Reengineering (SANER)}}. IEEE,
  \bibinfo{pages}{13--23}.
\newblock


\end{thebibliography}


\end{document}